# ALTERNATE MODELS TO DARK ENERGY

**Kenath Arun[1]**

Department of Physics, Christ University, Bengaluru-560029, Karnataka, India, &

Department of Physics, Christ Junior College, Bengaluru-560029, Karnataka, India

**S B Gudennavar**

Department of Physics, Christ University, Bengaluru-560029, Karnataka, India

**A Prasad**

Udaipur Solar Observatory, Physical Research Laboratory, Dewali, Bari Road, Udaipur 313001, India

**C Sivaram**

Indian Institute of Astrophysics, Bengaluru-560034, Karnataka, India

**Abstract:** One of the unresolved questions currently in cosmology is that of the non-linear accelerated expansion of the universe. This has been attributed to the so called Dark Energy (DE). The accelerated expansion of the universe is deduced from measurements of Type Ia supernovae. Here we propose alternate models to account for the Type Ia supernovae measurements without invoking dark energy.

**Keywords:** dark energy; dark matter; varying gravitational constant; varying proton mass

[1] Corresponding author:
e-mail: kenath.arun@cjc.christcollege.edu
Telephone: +91-80-4012 9292; Fax: +91-80- 4012 9222

# 1. Introduction

The detailed measurement of the mass density of the universe has revealed that about 70% of the energy density of the universe is unaccounted for. This appears to be connected to the independent observation of the non-linear accelerated expansion of the universe deduced from measurements of Type Ia supernovae (Riess et al., 1998; Perlmutter et al., 1999). Generally one would expect the rate of expansion to slow down, as once the universe started expanding, the combined gravity of all its constituents should pull it back, i.e. decelerate it. So the deceleration parameter, $q_0$, was expected to be a positive value.

A negative $q_0$ would imply an accelerating universe, with repulsive gravity and negative pressure. The measurements of Type Ia supernovae have revealed just that. By measuring their flux with redshift, $q_0$ is determined to be –0.55. This together with the fact that the universe is flat (from CMBR) and the total matter content, $\Omega_M \sim 0.3$, the rest of the matter in the universe, i.e., $\Omega_{DE} \sim 0.7$, must be in some exotic form which is dubbed Dark Energy (DE). All postulated forms of matter yield a positive deceleration parameter, except in the case of DE, hence this accelerated expansion is attributed to this dark energy (Peebles and Ratra, 2003; Sivaram, 2009).

# 2. Observational Evidence for Dark Energy

The luminosity distance of an object at a red shift of *z* is given by:

$$D_L = \frac{cz}{H_0}\left[1 + \frac{1-q_0}{2}z\right] \quad (1)$$

The flux reaching us from this distance is then given by (Weinberg, 1972):

$$l = \frac{L}{4\pi D_L^2} \quad (2)$$

where, *L* is the luminosity of the object.

In the case of a decelerating universe with say, $q_0 = 1/2$, we have the term in the bracket, $1 + \frac{1-q_0}{2}z = \frac{5}{4}$, and for an accelerating universe with say, $q_0 = -1/2$, we have $1 + \frac{1-q_0}{2}z = \frac{7}{4}$, at a redshift $z = 1$. So, a negative $q_0$ increases $D_L$. The flux is therefore smaller and the distant supernovae (SNe) appear fainter. The factor by which the SNe appears fainter is given by $\frac{7/4}{5/4}$.

Thus $D_L$ increases for $q_0 = -0.5$ by 7/5. So $l$ is smaller by $(7/5)^2 = 49/25 \approx 2$. The Supernova Cosmology Project reported the mass density and cosmological constant energy density of the universe based on the analysis of 42 Type Ia SNe. The magnitude-redshift data for these SNe, at redshifts between 0.18 and 0.83, indicate that the cosmological constant is non-zero and positive, with a confidence of P = 99%. For detailed analysis see (Perlmutter et al., 1999).

In the present work, we propose that this decrease in luminosity by a factor of 2, at $z = 1$, may be explained by some of the alternate models to DE without invoking a negative value of $q_0$. At $z = 2$, the factor by which $l$ is smaller is ~ 2.8.

## 3. Alternate Models

There are alternate models that tackle the dark energy problem through the extended theories of gravity, for example see (Corda, 2009). In the present work, some of the possible alternate models to dark energy that are considered are time varying $G$ and varying proton mass, and role of dark matter (DM) particles in white dwarfs lowering its mass limit and hence its inherent luminosity.

### 3.1. Time Varying G

One possible way to account for the fainter SNe is by a *time varying G*. The Chandrasekhar mass for a white dwarf is given by (Shapiro and Teukolsky, 1983):

$$M_{ch} = \left(\frac{\hbar c}{Gm_p^2}\right)^{3/2} m_p \qquad (3)$$

So for a higher $G$ at an earlier epoch, $M_{ch} \propto G^{-3/2}$ would be smaller. It was Dirac (1937), who first postulated that $G$ should decrease with time, to satisfy the so called Large Number Hypothesis (LNH). LNH notes that the ratio $e^2/Gm_p m_e$ (ratio of electrostatic to gravitational forces between a proton and electron) is just the same as the ratio of the Hubble age to a typical nuclear time scale, i.e., ~ $10^{40}$. So if it is not just a coincidence at the present epoch, $G$ should drop with time as the Hubble's constant, $H \propto t^{-1}$. Later on attempts to modify General Relativity (GR) to incorporate fully Mach's principle led to scalar-tensor theories such as the Brans-Dicke (B-D) theory. Here $G$ is a dynamical quantity, being a function of a scalar field, i.e.

$G \sim f(\phi)$, varying with time (and maybe space), in theories of type $G \sim \phi^{-1}$. Modern attempts to unify gravity with other interactions like superstring theory also require $G$ to vary with time. These theories also involve various scalar fields and in the limit (of low energy) lead to scalar tensor theories (of the B-D type).

$G$ here is also related to the size of the compacted extra dimensions which could shrink as usual space-time expands. This leads to variation of $G$ with time. $G$ could well have been 1% stronger $10^{10}$ years ago (a fractional change of $10^{-18}/s$). Some of the current limits are given by Mould and Uddin (2014) and Schlamminger et al. (2015). For example, if $G$ was 1% higher $M_{ch}$ would be 1.5% smaller. The temperature of a carbon-oxygen (C-O) white dwarf (most stars are expected to be C-O white dwarfs) core $T_c \propto M$ (from the virial theorem). Hence, $T_c$ would also be 1.5% smaller. Now the carbon detonation reaction is very sensitive to temperature, with the rate of C burning $\propto T^{30}$ (Caughlan and Fowler, 1988; Sivaram et al., 2014). Therefore a 1.5% change in the Chandrasekhar mass limit results in $(1.015)^{30} \approx 1.56$ change in the rate of C burning. This is nearly enough to make the type Ia SNe appear fainter. Figure 1 gives the variation of the SNe luminosity with the percentage change in $G$.

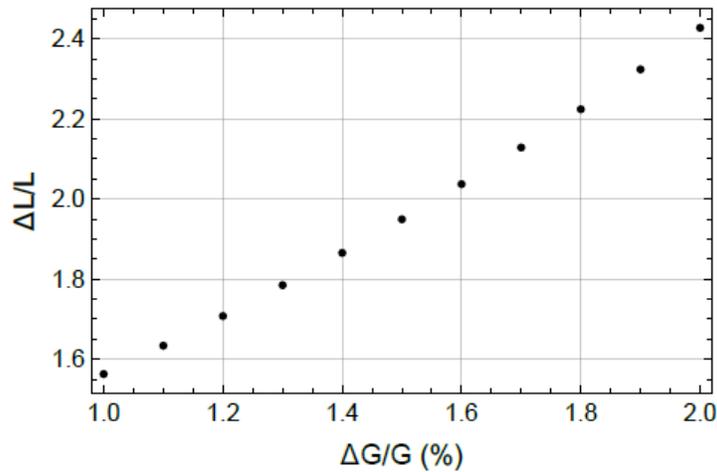

Figure 1 Change in luminosity with percentage change in G

The solar system tests of varying $G$ are of this order. But these tests may not unambiguously rule out a larger change in $G$ over the cosmological scales. It could have a general relation describing the time varying $G$ as:

$$G_z = G_0\left(\frac{t_0}{t}\right)^n \tag{4}$$

where $G_0$ is the value of $G$ at the present epoch $t_0$, where we have $\frac{t_0}{t} \cong (1+z)$, we thus obtain:

$$G_z = G_0(1+z)^n, \quad n < 1 \tag{5}$$

*3.2. Varying Proton Mass*

The above calculations are only suggestive that other alternatives could also be considered. LNH has also led to the suggestion that other constants could also vary, including the fine structure constant, $\alpha$, and the mass ratio $m_p/m_e$. We would next consider the latter case, i.e. variation of $m_p/m_e$ as yet another possibility, and study its effect on the $M_{ch}$. Current limits on the mass ratio $m_p/m_e$ is given in Dapra et al. (2016) and Hanneke et al. (2016). If the proton mass, $m_p$, and the ratio $m_p/m_e$ is larger, then $M_{ch}$ would be lower since it is $\propto 1/m_p^2$. A 1% change in the proton to electron mass ratio $m_p/m_e$, results in a 2% change in $M_{ch}$ and the detonation temperature would be 2% lower. The reaction rate, $(1.02)^{30} \approx 1.8$. This would be enough to account for the fainter SN. Figure 2 gives the variation of the SNe luminosity with the percentage change in proton mass.

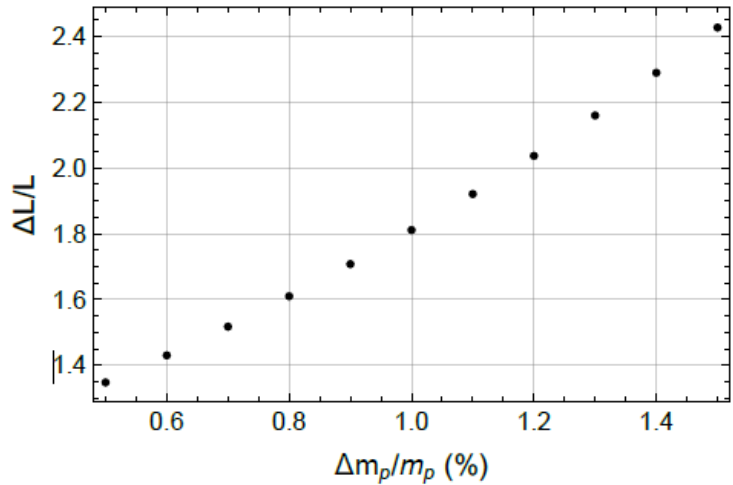

Figure 2 Change in luminosity with percentage change in proton-to-electron mass ratio

## 3.3. White Dwarf with Dark Matter Constituents

The mass limit for the White Dwarf (WD) varies with the constituent particle mass as, $M_{ch} \propto 1/m_p^2$. If the constituents are heavier, the WD mass would be less. As dark matter (DM) particle mass $m_D \gg m_p$ (generally $m_D$ is assumed to be around 10 GeV to 100 GeV, whereas $m_p \sim 1$ GeV), presence of even 1% of DM particles can lower $M_{ch}$ and make the supernova fainter. At earlier epochs, DM particles would have been denser by $(1+z)^3$, and hence it would be more likely for WD (as well as other dense stellar remnants) to accumulate DM particles. Also the stellar precursors, which led to the formation of WD at higher $z$, DM density would be $(1+z)^3$ more in the collapsing interstellar clouds (in early galaxies), so these stars could have more DM admixed with baryons. For example, if out of 100 particles (in WD), there are 10 DM particles (of 10 GeV) and 90 protons, this implies that instead of 100 GeV mass we now have 190 GeV. So, for 10% DM, the effective $m_p$ is increased by a factor of 1.9, and hence $M_{ch}$ is lower by a factor of $(1.9)^2 = 3.7$. Figure 3 gives the variation of $M_{ch}$ for different DM particle masses at different percentage of admixture.

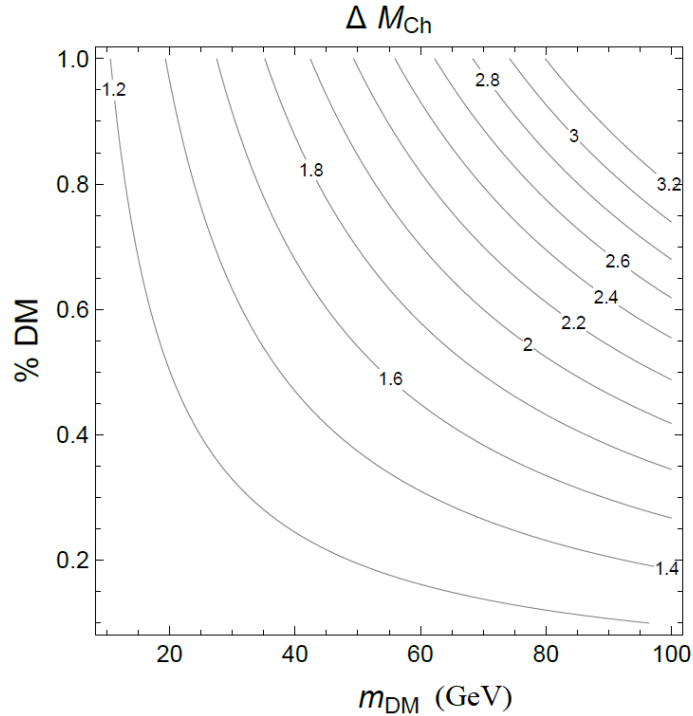

Figure 3 Change in $M_{ch}$ for varying DM particle mass with different percentage of accretion by the WD

About 3-4% DM in WD would be enough to lower $M_{ch}$ by about 1.9, which can fully account for the fainter SNe without requiring a negative $q_0$. If heavier DM particles are present, like 30 GeV or 60 GeV, even less than 1% admixture can lower $M_{ch}$ by a factor of 2. Table 1 gives the percentage of DM admixture required for different DM particle masses to account for a change in $M_{ch}$ by a factor of 2. So, this could well be another alternative to DE. This may also explain why so many neutron stars seem to have lower masses near to $M_{ch}$ rather than $> 2\ M_\odot$ as the neutron star mass limit could also be lowered due to admixture of DM particles.

| $m_D$ (GeV) | % DM |
|---|---|
| 20 | 2.180 |
| 30 | 1.428 |
| 40 | 1.062 |
| 50 | 0.845 |
| 60 | 0.702 |
| 70 | 0.600 |
| 80 | 0.524 |
| 90 | 0.465 |
| 100 | 0.418 |

Table 1 Percentage of DM required to change $M_{ch}$ by a factor of 2 for varying DM particle mass

*3.4. Accretion of Dark Matter Particles by White Dwarfs*

Another possible mechanism by which the inherent luminosity (and hence the flux) of the type Ia SNe can be reduced is by accretion of DM particles by WD pushing it over $M_{ch}$, or by the accretion of Jupiter or Neptune sized DM planets. In such cases what detonates would be a lower mass of the C-O white dwarf, hence making the SNe subluminous. DM concentrations in the ambient medium at earlier epochs would be $(1+z)^3$ times more. So early WD could accrete more DM and could be pushed over $M_{ch}$ to detonate but would be subluminous since the DM particles will not contribute to the luminosity. Accretion rate of the DM particles is given by:

$$\dot{M} = 4\pi r_{WD}^2 \rho_{DM} \text{v} = 3\times 10^{15}\ g/year \qquad (6)$$

At this rate of accretion, the WD would be pushed over $M_{ch}$ in about $10^{14}$ years. The concentration of DM is $10^5$ times higher in planetary systems than in interstellar space. Hence the accretion rate would be $3 \times 10^{20} \, g/year$. So in a period of $10^9$ years WD could accumulate DM to be pushed over $M_{ch}$. But C-O mass could be 50% lower, hence making these SNe subluminous. Even if WD accretes $10^{28}$ kg, it is sufficient to account for the subluminous SNe. This can be accreted over $10^9$ years or less. This will not affect the neutron stars (NS), since NS with $10^5$ lower surface area would accumulate less DM, hence there is not much change in the NS mass.

## 4. Summary and Conclusions

In this paper we have considered alternate explanations for the sub-luminous nature of the Type Ia SNe without invoking dark energy. We have looked at the possibility of varying fundamental constants as a possible means to explain this observation. As can be seen from Figure 1, a variation of *G* by even 1.5% (in ten billion years) is enough to account for the change in luminosity of the Type Ia SNe by a factor of 2. Similarly, as seen from Figure 2, even lesser variation of $m_p$ can account for this change in luminosity. We also consider the possibility of WD containing DM particles resulting in them being sub-luminous. Since the density of DM particles at earlier epochs was higher, we also consider the possibility of WD accreting these DM particles and as a result the luminosity of Type Ia SNe is reduced. As noted in Figure 3 and Table 1, an admixture of ~ 1% of DM particles is enough to account for the observed luminosity of the Type Ia SN. Considering that at early epoch the density of DM particles would be higher, this scenario is very much a possibility. So these possibilities imply that DE need not be inevitable to account for the dimness of distant Type Ia SN. In later works we plan to study the effects of the presence of DM particles in other stellar objects and the possible effects on their luminosities.